\documentclass[aps,pra,twocolumn,amsmath,amssymb,10pt]{revtex4-2}
\usepackage{amsmath,amssymb,graphicx,color,hyperref}
\usepackage{bm}
\usepackage{amsfonts}

\newcommand{\C}{\mathbb{C}}

\def\ZZ{\mathbb{Z}}

\newcommand{\bL}{\mathbf{L}}
\newcommand{\bM}{\mathbf{M}}

\newcommand{\bU}{\mathbf{U}}

\newcommand{\be}{\begin{equation}}
	\newcommand{\ee}{\end{equation}}
\newcommand{\bea}{\begin{eqnarray}}
	\newcommand{\eea}{\end{eqnarray}}

\newcommand{\ed}{\end{document}}

\newcommand{\bi}{\begin{itemize}}
\newcommand{\ei}{\end{itemize}}

\newcommand{\bce}{\begin{center}}
\newcommand{\ece}{\end{center}}

\newcommand{\RE}{{\rm Re}}
\newcommand{\IM}{{\rm Im}}

\begin{document}

\title{Non-Hermitian Floquet-Free Analytically Solvable Time Dependant Systems}

\author{Hamed Ghaemi-Dizicheh, Hamidreza Ramezani}
\email {hamidreza.ramezani@utrgv.edu}
\affiliation{Department of Physics and Astronomy, University of Texas Rio Grande Valley, Edinburg, Texas 78539, USA}

\begin{abstract}
The non-Hermitian models, which are symmetric under parity ($ P $) and time-reversal ($ T $) operators, are the cornerstone for the fabrication of new ultra-sensitive optoelectronic devices. However, providing the gain in such systems usually demands precise contorol of nonlinear processes, limiting their application. In this paper, to bypass this obstacle, we introduce a class of time-dependent non-Hermitian Hamiltonians (not necessarily Floquet) that can describe a two-level system with temporally modulated on-site potential and couplings. We show that implementing an appropriate non-Unitary gauge transformation converts the original system to an effective one with a balanced gain and loss. This will allow us to derive the evolution of states analytically. Our proposed class of Hamiltonians can be employed in different platforms such as electronic circuits, acoustics, and photonics to design structures with hidden $ PT $-symmetry potentially without imaginary onsite amplification and absorption mechanism to obtain an exceptional point.
	
\end{abstract}
\maketitle

\section{Introduction}

The non-Hermitian models in physics have become a vast research area in recent years, including condensed matter physics \cite{rui2019pt}, photonics \cite{longhi2010spectral,ding2015coalescence}, biophysics \cite{chou2011non}, and acoustic \cite{zhu2014p}. Among them, the non-Hermitian generalization of topological tight-binding systems introduces new phenomena absent in Hermitian ones, such as non-Hermitian skin effect \cite{yao2018edge}, distinct transport effects \cite{ghaemiPRA2021}, relocation of topological edge states \cite{hamid-2022}, and noise-resilient  \cite{Tuxbury_2022}, to name a few.\\ A well-known approach to making a tight-binding model non-Hermitian is introducing complex onsite potential, playing the role of gain (loss) in photonic models \cite{schomerus2013topologically}. One can also stagger inter (intra) cell hopping amplitudes and develop non-reciprocity through the system \cite{lee2016anomalous,yao2018edge,lieu2018topological}.\\
Among non-Hermitian hamiltonians with gain and loss, the ones with balanced amplification and absorption are known as parity time-reversal ($PT$) symmetric hamiltonians and attracted more intention. $PT$-symmetric systems have been studied in both continuous and discrete models \cite{feng2017non,el2018non} and extensively investigated theoretically and experimentally in photonic lattices \cite{el2007theory,makris2008beam,ruter2010observation,regensburger2012parity}, microring resonator \cite{peng2014loss,hodaei2014parity,peng2014parity}, and electrical circuit \cite{schindler2011experimental,chitsazi2017experimental,leon2018observation}.\\
In $PT$-symmetric non-Hermitian systems, specific singular points are known as exceptional points (EPs) where at least two or more eigenvalues and corresponding eigenvectors of the hamiltonians coalesce. In recent years, the EPs have led to many interesting phenomena observed in different physical systems \cite{berry2004physics,bender2007making,rotter2009non,moiseyev2011non,heiss2012physics,cao2015dielectric}.\\
Indeed, EPs separate the broken $ PT $-symmetric regions, in which the eigenvalues of the non-Hermitian Hamiltonian become complex, from the unbroken $ PT $-symmetric phase with real-valued eigenvalues.
Many nontrivial features such as unidirectional invisibility \cite{lin2011unidirectional}, invisible sensor \cite{fleury2015invisible}, single-mode lasing \cite{feng2014single}, sensitive readout \cite{dong2019sensitive} and robust wireless power transfer \cite{assawaworrarit2020robust} can be investigated by creating and developing $ PT $-symmetric structure. However, from a practical point of view, constructing delicate balanced gain and loss is a rigorous requirement that hinders the possibility of implementing $ PT $-symmetry and studying its interesting features in specifically quantum systems.\\In addition, fundamental obstacles such as gain-induced noise and instabilities arise motivation to look for a different route of achieving $ PT $-symmetry without gain. The gain-free models with a hidden $ PT $-symmetry capture various research interests \cite{li2020virtual,Li2020timemodulatedpt,guo2009observation,ornigotti2014quasi,feng2013experimental,peng2014loss,liu2019observation}.\\ The key approaches to design a system benefit hidden $ PT $-symmetry, are based on nonlinear phenomena \cite{Yue2019PTnonlinear} or by making use of a temporally modulated coupling. In the latter method, a unitary gauge transformation leads to an effective system whose Hamiltonian is invariant under parity and time reversal. A well-known platform to explore such a gauge transformation is an electronic system. For example, a transient $ PT $-symmetry can be triggered by the switching on and off of electronic devices \cite{yang2022observation}.\\
A larger class of Hamiltonian with hidden $ PT $-symmetry can be developed and studied by releasing the unitarity condition on the transformation to reach the effective Hamiltonian. This approach shapes the main idea of this paper.\\ In this study, we introduce a general class of time-dependent two-level non-Hermitian Hamiltonians with a hidden $ PT $-symmetry which can be revealed by implementing a non-unitary gauge transformation. The main aspect of this approach is its generality. Indeed, our proposed Hamiltonian is not \textit{necessarily floquet} which makes it applicable to the vast range of time-dependent two-level systems. In addition, we formulated our two-level Hamiltonian with an arbitrary onsite potential. In the case of onsite potential with zero imaginary part, the system can be transformed into the effective one with balanced gain and loss. Therefore our proposed method can bypass the usual difficulties in realizing amplification and absorption mechanism. The two-level Hamiltonian investigated in this letter can model acoustics, electronics, and photonics systems whose couplings are time or coordinate dependant. \\Besides, we show that the analysis of state evolution described by this class of Hamiltonian can be analytically obtained regardless of adiabatic condition. The exact solutions are helpful to predict and manipulate the system's behavior precisely when it encircles EPs \cite{hassan2017dynamically}.\\ 
We organize the paper as follows. In section \ref{section1}, we introduce the Hamiltonian, non-unitary time dependant transformation and manifest the $ PT $-symmetry in the model. In section \ref{section 2}, we apply our method to an electronic system to illustrate different $ PT $-symmetric phases. The system's time evolution is studied analytically in the appendix.
\section{Time-dependant model}\label{section1}
\begin{figure*}[t]
	\begin{center}
		\includegraphics[scale=0.55]{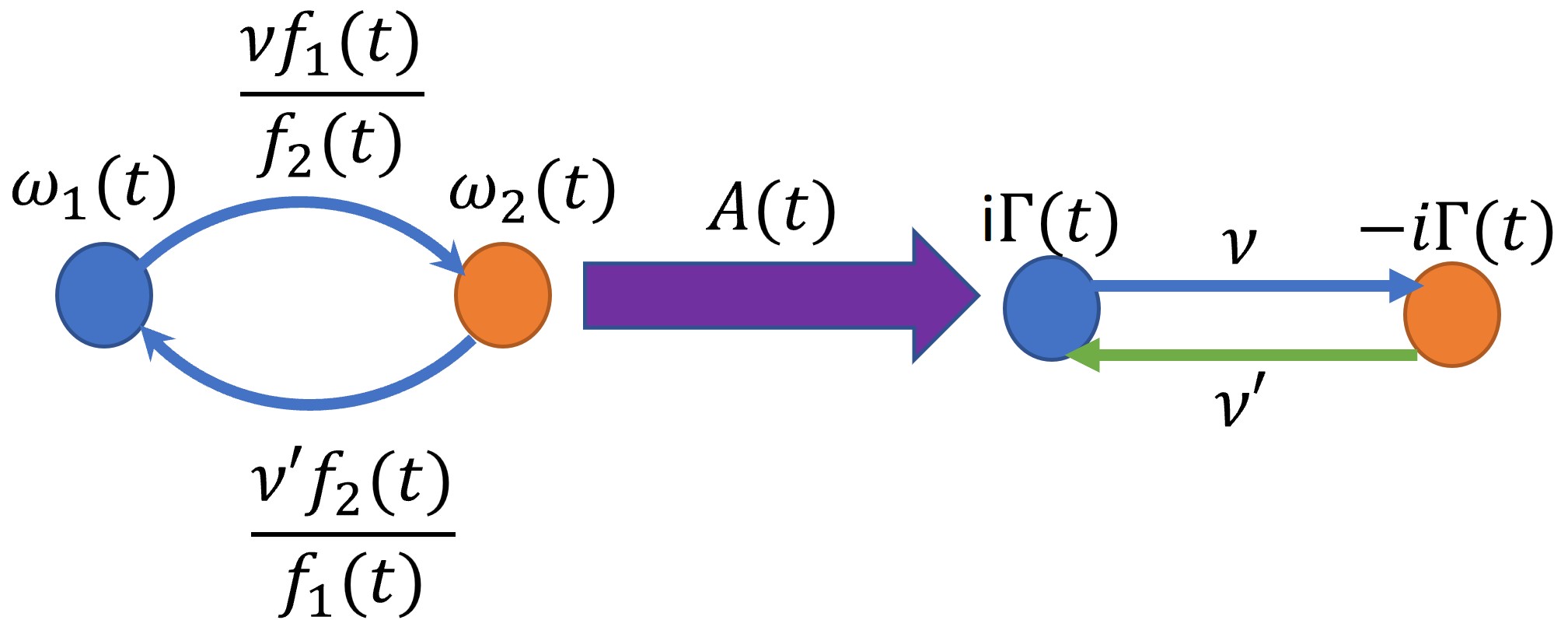}
		\caption{A two-level model which is described by the Hamiltonian \ref{time-dependant Hamiltonian} transforms to a model with balanced gain and loss by a time-dependent transformation $ A(t) $. This system can describe two optical resonators coupled to each other nonreciprocally in which their coupling is modulated in time. In this model, the resonator 1 or 2 provides dynamical gain (loss) for $ \IM(\omega_{1,2})<0 $ ( $ \IM(\omega_{1,2})>0 $).} \label{fig:setup}
	\end{center}
\end{figure*}
Let us consider a class of two-level systems described by the following time-dependent Hamiltonian
\begin{align}
	H(t)=\left( \begin{array}{cc}
		\omega_1(t)&\dfrac{\nu f_1(t)}{f_2(t)}  \\
		\dfrac{\nu' f_2(t)}{f_1(t)} &\omega_2(t) 
	\end{array}\right), \label{time-dependant Hamiltonian}
\end{align}
where $ \nu,\nu'$ are real-valued constants representing time-independent coupling, $ \omega_1(t)$ and  $ \omega_2(t)$ are complex-valued time-dependent onsite potentials, and we introduce nonzero finite-range timely modulated couplings $ f_{1,2}(t) $ which are generally linearly independent and in an electronic system they represents time-dependent capacitor and inductance \cite{yang2022observation}. In the following, we will show how this particular type of coupling, combined with an appropriate non-unitary gauge transformation, results in an effective Hamiltonian with balanced gain and loss. The time-dependent Schr\"{o}dinger equation reads $  $
\be
i\partial_t \Psi(t) =H(t)\Psi(t),\label{basic SE}
\ee 
where $ \Psi(t) =(\psi_1(t),\psi_2(t))^T $ is the state vector such that for a given initial value $ \Psi(t_0) $, the dynamical evolution of modes $ \psi_{1,2} $ is given via linear differential equation (\ref{basic SE}).\\
We introduce time-dependent transformation $\Psi(t) =A(t)\chi(t)$, which converts the Hamiltonian (\ref{time-dependant Hamiltonian}) into the effective Hamiltonian
\be
H\rightarrow H_{\text{eff}}(t):=A^{-1}HA-i\partial_t \ln A,\label{Transformed Hamiltonian}
\ee
where $ A $ is an invertible ($ \det A\neq 0$) matrix. For a closed two-level system with a reciprocal coupling between their sites, the Hamiltonian is given by a Hermitian matrix (i.e., $ H=H^{\dagger} $). For such a system, the transformation $ A $ needs to be given by a unitary matrix to conserve the norm of state (i.e., $\|\Psi(t)\|= \|\chi(t)\|$), in this case, the two terms on the right side of (\ref{Transformed Hamiltonian}) are also Hermitian. However, in our study, we do not restrict our approach to the similarity condition and consider a general time-dependent transformation. Specifically, by choosing a suitable form of time-dependent transformation $ A(t) $, one can reduce the Hamiltonian $ H(t) $ to a traceless matrix (i.e., Tr($ H_{\text{eff}})=0$) which describes the effective system with balanced gain and loss. This can be achieved by
\begin{align}
	A(t)=\exp\left( -i\int^{t}\Omega_-(t')dt'\right) \left( \begin{array}{cc}
		\sqrt{\frac{f_1(t)}{f_2(t)}}&0  \\
		0 &\sqrt{\frac{f_2(t)}{f_1(t)}} 
	\end{array}\right), \label{time-dependant gauge transformation}
\end{align}
where, $ \Omega_\pm(t):=(\omega_1(t)\pm\omega_2(t))/2 $. By substituting $ A(t) $ into (\ref{Transformed Hamiltonian}), the off-diagonal terms become time-independent, and the effective Hamiltonian turns to
\begin{align}
	H_{\text{eff}}(t)=-i\Gamma\sigma_{z}+\nu\sigma_{+}+\nu'\sigma_{-}, \label{effective Hamiltonian}
\end{align}
where
\be
\Gamma(t):=i\Omega_{-}+\frac{1}{2}\partial_t\ln\frac{f_1}{f_2}, \label{Gamma}
\ee
is an complex-valued effective onsite potential and $ \sigma_{\pm}:(\sigma_x\pm i\sigma_y)/2 $ are given in terms of Pauli matrices $ \sigma_{x,y,z} $ (see Fig. \ref{fig:setup} for the schematic of the transformation). For the effective Hamiltonian (\ref{effective Hamiltonian}), Schr\"{o}dinger equation then reads as
\be
i\partial_t\chi(t) =H_{\text{eff}}(t)\chi(t),\label{Effective SE}
\ee 
where the transformed vector state is now introduced with $ \chi(t)=(\zeta_-(t),\zeta_+(t))^T $.\\
By choosing a specific form of time transformation $ A $ given by (\ref{time-dependant gauge transformation}), the Schr\"{o}dinger equation (\ref{Effective SE}) can be recasted into the following second-order differential equation for $ \zeta_{\mp} $
\be
\dfrac{d^2\zeta_{\mp}}{d\tau^2}\pm\left[\partial_{\tau}\Gamma(\tau)\mp i\Gamma(\tau)^2\pm1 \right]\zeta_{\mp}=0. \label{decopled differential equation}
\ee
In driving the above differential equations, we scaled variables such as
\be
\left(\dfrac{\Gamma(t)}{\sqrt{\nu'\nu}},\sqrt{\nu'\nu} t\right) \rightarrow (\Gamma(\tau),\tau).
\ee
If we suppose $ \omega_2(t)=0 $, then, the vector state $  \Psi(t) $ is given by
\be
(\psi_1,\psi_2)=e^{\int^\tau\Gamma(\tau')d\tau'}(f_1f_2^{-1}\zeta_{-},\zeta_+)\label{exact solution of psi}.
\ee
For the nonzero $ \omega_2(t)\neq 0 $, the evolution of $ \psi_{1,2} $ are given by the above equation by replacing $ \Gamma(\tau)\rightarrow \omega_2(\tau)-\Gamma(\tau) $.\\
In the appendix, by considering three different forms for the effective onsite potential $ \Gamma(t) $, we show that the evolution of the system governed by (\ref{Effective SE}) can be obtained analytically. From there, implementation of an inverse time transformation leads to the dynamics of the original system given by (\ref{time-dependant Hamiltonian}).\\
To conclude the section, we are looking through the eigenbasis and eigenvalues of the original Hamiltonian (\ref{time-dependant Hamiltonian}) and its effective counterpart (\ref{effective Hamiltonian}). For the effective system, the time-dependent eigenvalues are given by
\be
\bar{\lambda}_{\mp}=\mp\sqrt{\nu\nu'-\Gamma^2(t)}.
\ee
In the case of static $ PT$-symmetric where $ \Gamma $ is time independent and $ \nu=\nu' $, the system is in $ PT $ unbroken (broken) phase for $ \vert\nu\vert>\Gamma $ ($ \vert\nu\vert<\Gamma $), where $ \vert\nu\vert=\Gamma  $ is an exceptional point. In the case of the periodic Hamiltonian, then the $ PT $ (un)broken phase can be found by looking at the exceptional points of the Floquet Hamiltonian.\\ For the general time-dependant $ \Gamma(t) $, as a consequence of non-Hermicity, the eigenbasis of the system is not orthogonal, but one can define biorthogonal eigenbasis by using its right and left eigenvectors which are defined via
\begin{align}
	&H_{\text{eff}} \chi_\mp^r=\bar{\lambda}_\mp \chi_\mp^r ,&&H_{\text{eff}}^{T} \chi_\mp^l=\bar{\lambda}_\mp \chi_\mp^l,
\end{align}
where $ T $ stands for the transpose of the matrix. However, for the reciprocal system where $ \nu=\nu' $ and the invariant effective hamiltonian under transpose (i.e., $ H_{\text{eff}}^T=H_{\text{eff}} $), one can find the similar right and left eigenbasis and construct a biorthogonal pair just by picking up one of them. In this case, then the eigenbasis is given as \cite{milburn2015general}
\bea
&&\chi_-^r=\chi_-^l=(-\sin\frac{\bar{g}}{2},\cos\frac{\bar{g}}{2})^T,\nonumber\\
\nonumber\\
&&\chi_+^r=\chi_+^l=(\cos\frac{\bar{g}}{2},\sin\frac{\bar{g}}{2})^T,
\eea
with $ \bar{g} $ such that $ \cot\bar{g}=i\Gamma/\nu $. On the other side, for the original Hamiltonian, the eigenvalues are
\be
\lambda_{\mp}=\Omega_+\mp\sqrt{\Omega_-^2+\nu\nu'},
\ee
and the right eigenbasis is given by
\bea
&&\Psi_-^r=(-f_1(t)\sin\frac{g}{2},f_2(t)\cos\frac{g}{2})^T,\nonumber\\
\nonumber\\
&&\Psi_+^r=(f_1(t)\cos\frac{g}{2},f_2(t)\sin\frac{g}{2})^T,
\eea
with $ g $ given by $ \cot g=\Omega_-/\nu $. The relation between $ g $ and $ \bar{g} $ can be found such as
\be
\cot(\bar{g})=\cot(g)-\frac{iW(f_{1}f_{2})}{2\nu f_{1}f_{2}},\label{cot}
\ee
where $W(f_{1}f_{2})$ is the Wronskian of two differentiable functions $f_{1}$ and $f_{2}$. From Eq. (\ref{cot}), one can find $ g $ in terms of $ \bar{g} $ from
\begin{align}
	&g=i\coth^{-1}(\frac{1}{2\nu}\partial_t\ln \frac{f_1}{f_2}+i\cot\bar{g})+\pi k,&&k\in\ZZ.
\end{align}
For the original system, if $ \Omega_-=\pm i\nu $, it reaches exceptional points (EPs). By following the same calculations for the effective system given by the Hamiltonian $ H_{\text{eff}} $, the EPs exist if
\be
\Omega_-=\mp i\nu+\frac{i}{2}\partial_t\ln(f_1f_2).\label{effective EPS}
\ee 
To study the dynamic of EPs under time-dependent transformation, we furthermore define the following complex parameter
\be
B(t):=\Omega_-(t)/\nu=\cot g.
\ee
In regarding the original Hamiltonian (\ref{time-dependant Hamiltonian}), the complex parameter $ B $ takes a complex-valued number in the parameter space ($ \RE(B) $-$ \IM(B) $), which are
\be
B^{P}=\pm i.
\ee
Under the transformation $ A(t) $, the loci of EPs for the effective system in its parameter space ($ \RE(\bar{B}) $-$ \IM(\bar{B}) $) are given by $ \bar{B}^{P}=\pm i $ where 
\be
\bar{B}(t):=\dfrac{i\Gamma(t)}{\nu}=\cot\bar{g}.
\ee
Tracking the loci of EPs for the effective system in the $ B $-parameter space shows that they are no longer constant and given in time as 
\be
\bar{B}^{P}=B^{P}-ib(t),\label{transformed B-parameter}
\ee
where $ b(t):=\frac{\partial_t\ln \frac{f_1}{f_2}}{2\nu}=b_r(t)+ib_i(t) $. To derive (\ref{transformed B-parameter}), we put $ \cot g=\cot\bar{g}=i $ in Eq. (\ref{cot}). It shows that, in the $ B $-parameter space, the real (imaginary) coordinate at $ t=t_1 $ of EPs shift by the factor of $ -b_i(t_1) $($ b_r(t_1) $).\\
In the following section, by making use of an electronic platform, we illustrate the hidden $PT$-symmetric in the system whose dynamics are given by the Hamiltonian (\ref{time-dependant Hamiltonian}).
\section{Application: $ PT $ symmetry in an electronic system}\label{section 2}
\begin{figure}
	\begin{center}
		\includegraphics[scale=.45]{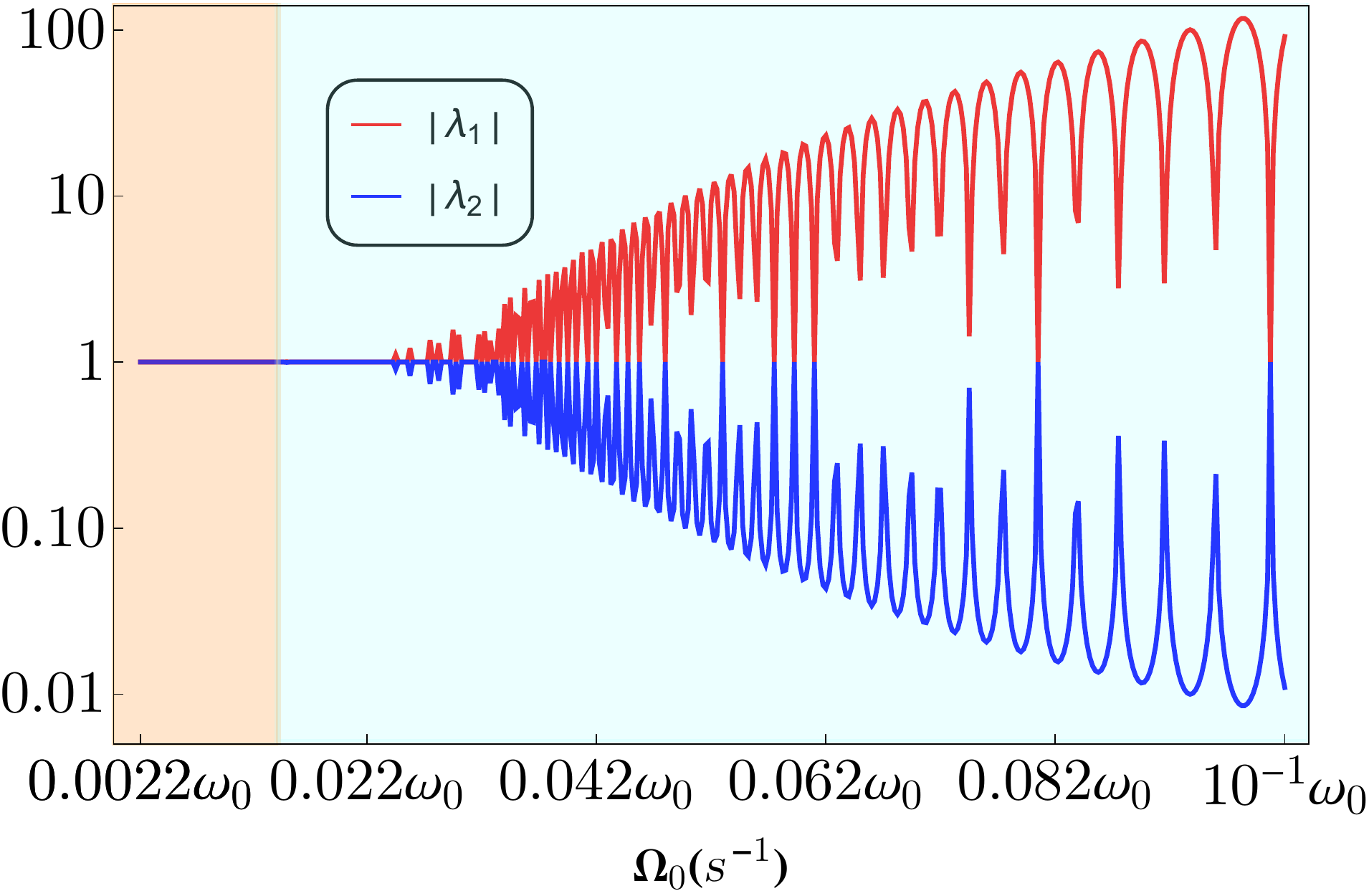}
		\caption{The magnitude of the eigenvalues $ \lambda_k $ belong to the time evolution operator $ U_F $ versus $ \Omega_0 $. Orange and blue regions represent the $ PT $ unbroken and broken phases of the system, respectively.}
		\label{Floquet1}
	\end{center}
\end{figure}

We implement our approach to design a $PT$-symmetric model using an electronic platform consisting of an RLC circuit. By following the Kirchhoff's laws, one can find the coupled equation for the voltage in capacitor $ V(t) $ and the current across the inductor $ I(t) $:
\bea
&&\partial_t V(t)=-\frac{I(t)}{C},\\
&&\partial_t I(t)=\frac{V(t)}{L}+\frac{R}{L}I(t).
\eea
By taking $ \Psi(t)=(V(t),I(t))^{T} $, the dynamic of $ RLC $ circuit can be given with $ H(t) $
\begin{align}
	H(t)=i\left( \begin{array}{cc}
		0&-1/C  \\
		1/L &R/L 
	\end{array}\right). \label{time-dependant Hamiltonian for RLC}
\end{align}
For a time-dependent resistance $ R(t) $, capacitance $C(t)$ and inductance $L(t) $, the above Hamiltonian-like matrix can be written in the form of the Hamiltonian (\ref{time-dependant Hamiltonian}) if we set
\begin{align}
	&L(t)=L_0 f(t),&C(t)=C_0 f^{-1}(t),&&R(t)=g(t).
\end{align}
\begin{figure}
	\begin{center}
		\includegraphics[width=0.95\linewidth, angle=0]{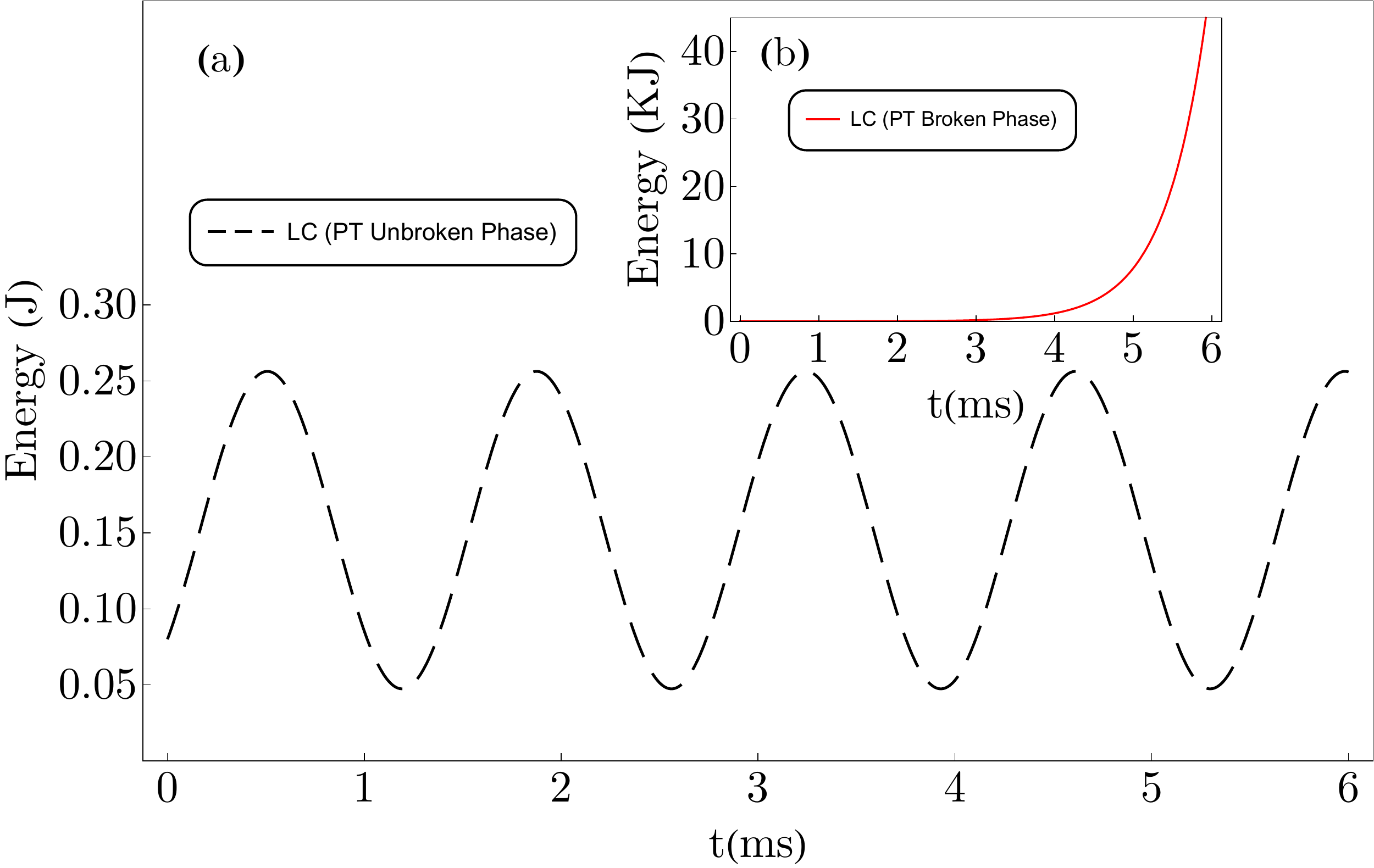}\hspace{0.5cm}
		\includegraphics[width=0.95\linewidth, angle=0]{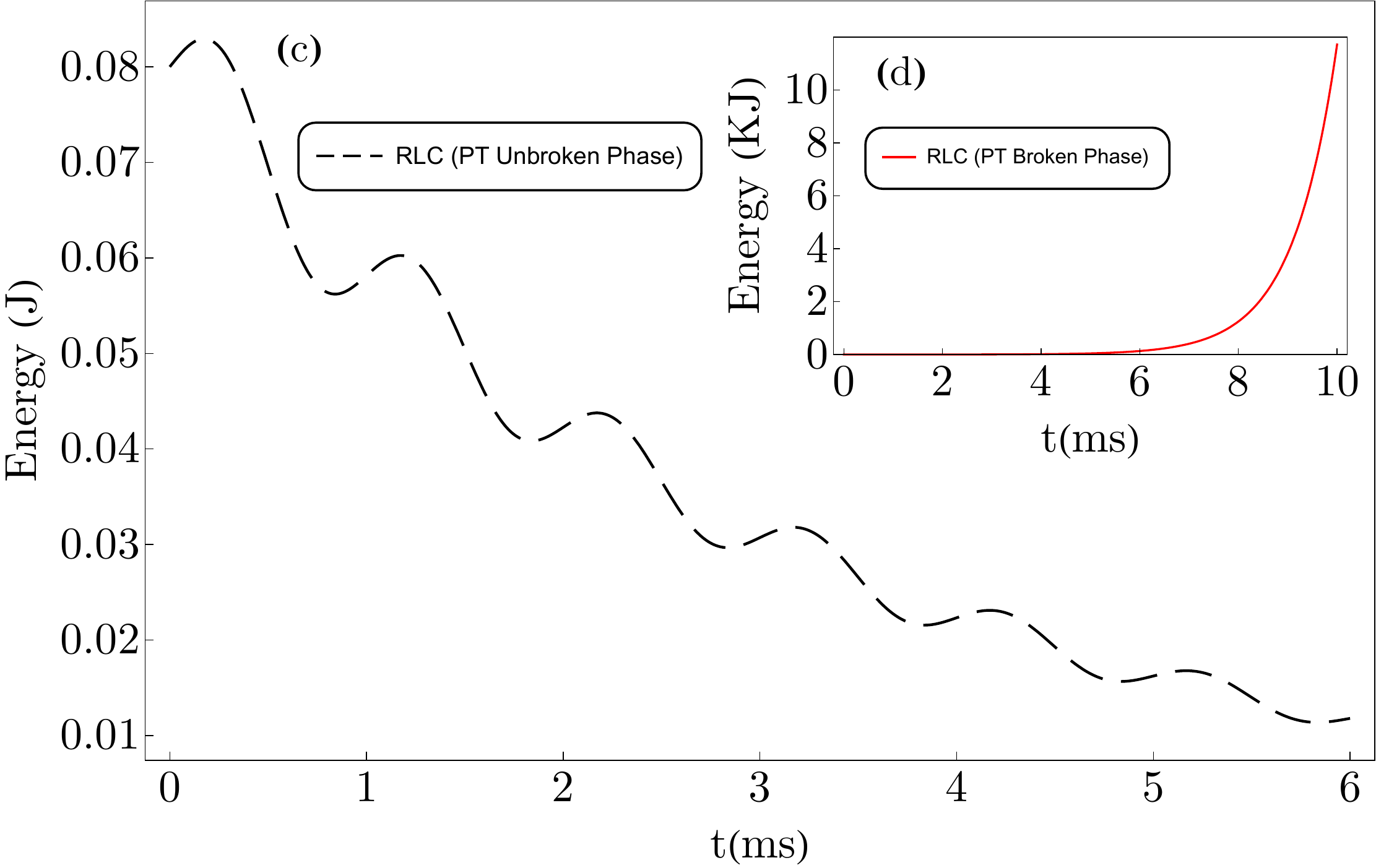}
		\caption{Time evolution of the circuit’s energy for LC circuit (a,b) and RLC circuit (c,d). In $ \gamma<\omega_0 $ ($ \gamma<2\omega_0 $) regime, the system is in the unbroken $ PT$-symmetric phase, and the energy oscillates. This oscillation is depicted in plots (a) and (c), where, in the latter, the resistance causes the damping oscillation. In the broken phase, the energy amplifies for both electronic systems.}
		\label{PT RLC}
	\end{center}
\end{figure}
Then in terms of our denotations, we have
\bea
&&f_1(t):=f(t),~~~f_2(t):=1,\\
&&\omega_1(t):=0,~~~~~~\omega_2(t)=ig(t)/L_{0}f(t),\\
&&\nu:=-i/C_0,~~~~\nu':=i/L_0.
\eea
If we modulate the resistance such that $ g(t)=L_0\partial_t f $, then the effective Hamiltonian (\ref{effective Hamiltonian}) reduces to
\be
H_{\text{eff}}=i\partial_t\ln f\sigma_z+\omega_0\sigma_y,
\ee
where $ \omega_0=1/\sqrt{L_0C_0} $. To get the above relation, we modified the time transformation (\ref{time-dependant gauge transformation}) such that $ A(t)\rightarrow \text{diag}(C_0^{-1/2},L_0^{-1/2})A(t)$ \cite{footnote1}. We first study the time-independent, effective Hamiltonian corresponding to case (a) in the appendix. This can be achieved by setting $ f(t)=e^{-\gamma t} $. Then, the electronic $ RLC $ circuit stimulate a static $ PT $-symmetric system whose Hamiltonian is given by $ H_{\text{eff}}^{PT}=i\gamma \sigma_z+\omega_0\sigma_y $. The eigenvalues of this Hamiltonian are given by $ \lambda_{\pm}=\pm\sqrt{\omega_0^2-\gamma^2} $. From these eigenvalues, we can find the $ PT $-symmetry breaking threshold where the system is referred to as being in a $ PT $-broken phase for $ \gamma>\omega_0 $. Fig. \ref{PT RLC} demonstrates the time evolution of the circuit's energy $ U_L+U_C $ for both LC ($ R(t)=0 $) and RLC circuit where
\begin{align}
	&U_L:=\frac{1}{2}L(t)I^2(t),&&U_C:=\frac{1}{2}C(t)V^2(t).
\end{align}
In the regime $ \gamma<\omega_0 $ ($ \gamma<2\omega_0 $), which defines unbroken $ PT $-symmetric phase, one can see from the diagram that the energy of the RLC (LC) system oscillates while in the broken phase, the energy of both systems amplifies. Note that in the RLC system, the oscillation is damping by the factor of $ e^{-R/2L} $ because of the resistance.\\
In the second step, we consider a periodic time-dependent $ f(t)=\epsilon_1\cos(\Omega_0 t)+\epsilon_2 $ with the period $ T=2\pi/\Omega_0 $. In this case, the effective Hamiltonian becomes periodic, and therefore its dynamics are determined by the one-period, non-unitary time evolution operator obtained from the following time-ordered product \cite{shirley1965solution},
\be
U(T)=\mathbb{T}e^{-i\int_0^TH_{\text{eff}}(\tau')d\tau'}\equiv e^{-iTH_F},\label{time evolution}
\ee
where the time-independent operator $ H_F $ is the so-called Floquet Hamiltonian in which their eigenvalues and eigenstates are called quasi-energies, and quasi-states \cite{barone1977floquet}. The corresponding Floquet Hamiltonian $ H_F $ can determine the $ PT $ broken and unbroken phases of the system. To show this, we consider the eigenvalues of Floquet Hamiltonian $ \varepsilon_{1,2} $ and the time evolution $ \lambda_{1,2} $ that related to each other through $\lambda_{1,2}=e^{-iT\varepsilon_{1,2}} $. The quasi-energies $ \varepsilon_{1,2} $ take real value in the phase where the system is invariant under parity and time-reversal transformation. In this phase, the eigenvalues of the evolution operator $ U_F $ satisfies $ \vert\lambda_1\vert $=$ \vert\lambda_2\vert=1 $ (see Fig. \ref{Floquet1}). The energy of the time-dependent RLC circuit oscillates in this phase while it is damping when the Floquet Hamiltonian enters the broken $PT$ phase.\\ We note that for time-dependent coupling in the form of the step function, the $ PT $ (un)broken phases have been studied in \cite{quiroz2021parity} for the LC circuit. However, our approach can encompass a general class of time-dependent systems with hidden $ PT $-symmetric phase.
\section{Conclusion}
In summary, in this paper, we propose a class of time-dependent Hamiltonian describing a two-level system with the hidden $ PT $-symmetry. The hidden symmetric can be revealed by implementing a specific non-Unitary gauge transformation which transforms the original Hamiltonian with temporally modulated couplings and onsite potentials to an effective one with balanced gain and loss. The dynamics of the effective system can be analytically extracted. Time-dependent systems are not usually analytically solvable. Thus, using our method we can propose a class of systems that one can analytically drive the condition for the existence of exceptional points and thus use it for determining other applications of exceptional points including sensing. The approach of this paper can provide a protocol to design and develop a different type of system with a $ PT $-symmetric characteristic without encountering the obstacles of injecting gain. We demonstrated how our method can be applied to electronic devices. However, photonics models may have a potential application. The main experimental challenge in developing a photonics application is designing a system with time-varying couplings and onsite potentials. Microwave Smoothly deformed metallic waveguides in the microwave and helical waveguides in optics have the potential to be used in the design of driven photonics systems. In such a system, the deformed waveguide is extended along the $x$-axis, allowing us to achieve the spatially modulated couplings \cite{Rechtsman_2013, doppler2016dynamically}, where $x$ now represents time in our approach. 

\begin{acknowledgments}
H. R. acknowledge the support by the Army Research Office Grant No. W911NF-20-1-0276 and NSF Grant No. PHY-2012172 and OMA-2231387. 
\end{acknowledgments}
\appendix
\setcounter{secnumdepth}{0}
\section{Appendix: Analytic Solution}\label{appendixa}
In this appendix, we study the analytic solutions of (\ref{effective Hamiltonian}) by considering different forms for $ \Gamma(t) $.\\
\textbf{Case (a).} We choose the effective potential such that the original Hamiltonian (\ref{time-dependant Hamiltonian}) transforms to the time-independent effective Hamiltonian. It corresponds to set $ \Gamma(\tau)=\gamma $, where $ \gamma\in \C $ is a real-valued constant. Since, in general, the time-dependent transformation $ A(t) $ in (\ref{time-dependant gauge transformation}) is a non-similarity transformation, then the eigenvalues of the system do not remain unchanged. Regarding this, we can design a system with linearly, time-modulated Hamiltonian such as $ H(t) $, which effectively acts as a time-independent system whose eigenenergies are constant. One practical way to apply this protocol is to make use of an electronic platform. Regarding this choice of effective on-site potential, from differential equation (\ref{decopled differential equation}), one can find the following solution for $ \zeta_\mp $:
\be
\zeta_{\mp}=c_1^{(\mp)}e^{i\bar{\gamma}\tau}+c_2^{(\mp)}e^{-i\bar{\gamma}\tau},
\ee
where $\bar{\gamma}:=\sqrt{1-\gamma^2} $, and $ c_j^{(\mp)} $ are constants given by the initial values of $ \zeta_{-}(0):=a $ and $ \zeta_{+}(0):=b $ such as
\bea
&&c_1^{(-)}=\dfrac{\bar{\gamma}(1+i)a+i\nu b}{2\bar{\gamma}},~~c_2^{(-)}=\dfrac{\bar{\gamma}(1-i)a-i\nu b}{2\bar{\gamma}},\nonumber\\
&&c_1^{(+)}=\dfrac{i\nu'a+\bar{\gamma}(1-i)b}{2\bar{\gamma}},~~c_2^{(+)}=\dfrac{-i\nu'a+\bar{\gamma}(1+i)b}{2\bar{\gamma}}.\nonumber\\
\eea
If $ \omega_2(t)=0 $, from (\ref{exact solution of psi}), the vector state $ \Psi(\tau)$ is given by
\be
\Psi(\tau)=e^{-\gamma \tau}(f_1f_2^{-1}\zeta_{-},\zeta_{+})^T.
\ee
Now let $ \gamma:=(\gamma_1+\gamma_2)/2\sqrt{\nu\nu'} $, then in light of the definition of $ \Omega(\tau) $ and $ \Gamma(\tau) $, one can find the relation between the one-site potential $ \omega_j(t) $ and coupling $ f_j(t) $ such that:
\begin{align}
	&\omega_1(t)-\dfrac{if'_1(t)}{f_1(t)}=-i\gamma_1,&\omega_2(t)-\dfrac{if'_2(t)}{f_2(t)}=-i\gamma_2.\label{Eq. for one-site potential}
\end{align}
These lead to the following equation
\be
\dfrac{f_1(t)}{f_2(t)}=f_0 u(t) e^{i\int[(\omega_1(t)-\omega_2(t))]dt},\label{Eq. for coupling}
\ee
where $ f_0:=\dfrac{f_1(0)}{f_2(0)} $ and $ u(t):=e^{2(\gamma_1-\gamma_2) t} $.
\begin{figure}
	\begin{center}
		\includegraphics[scale=.4]{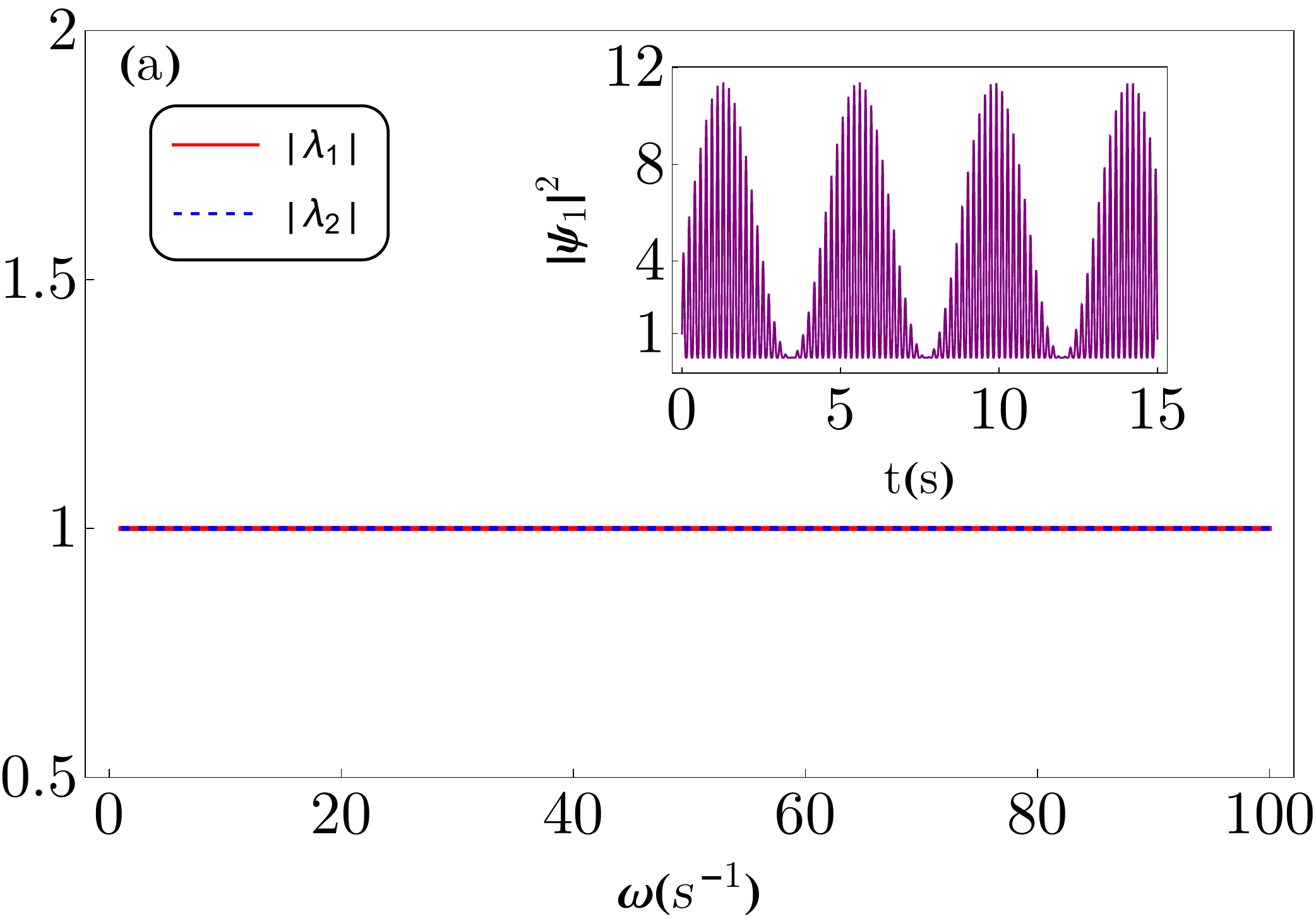}\hspace{0.5cm}
		\includegraphics[scale=.4]{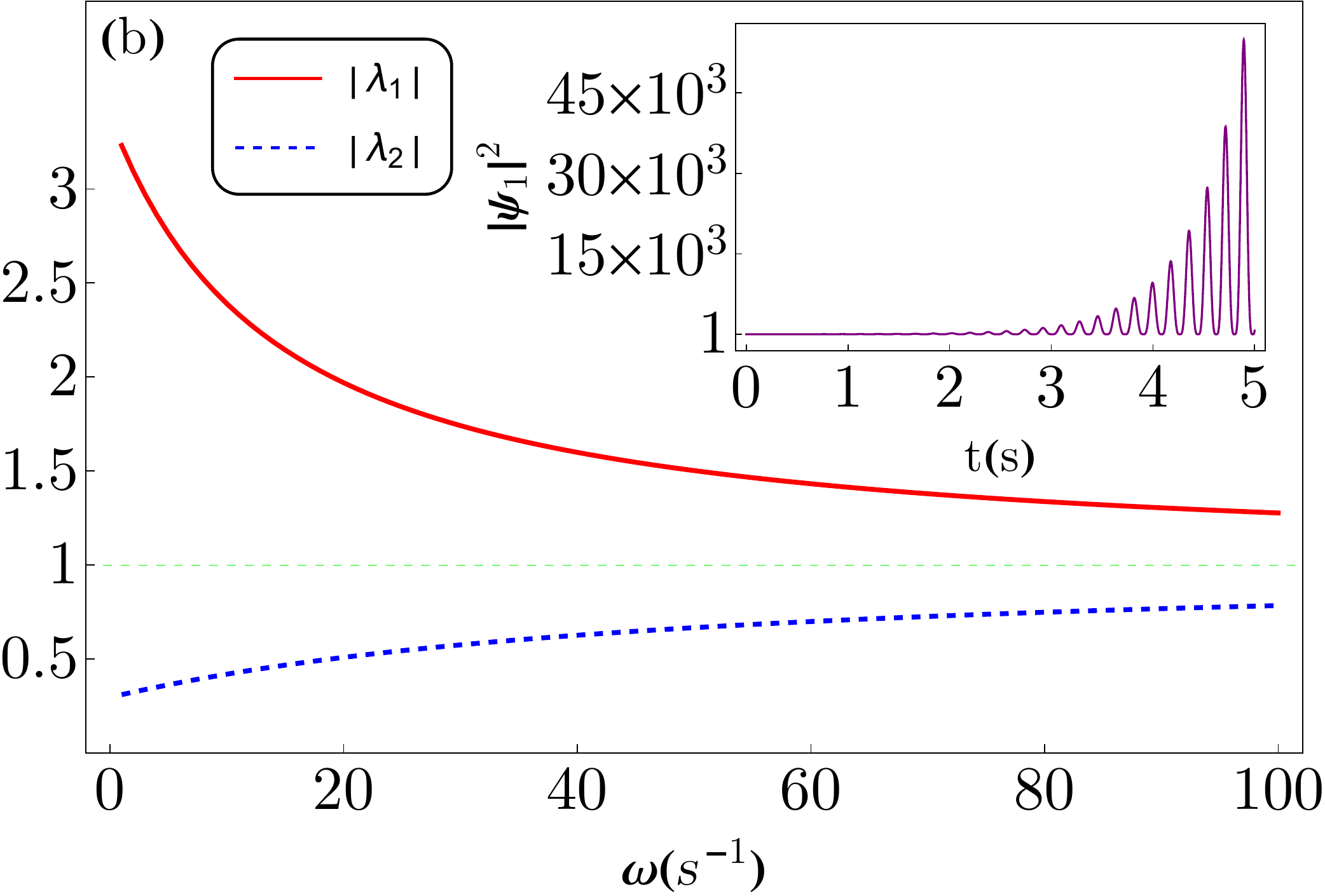}
		\caption{The quasienergies and the mode density for the toy model given with the parameter given in relations (\ref{toyparameter1}) and (\ref{toyparameter2}) in the unbroken (a) and broken (b) $ PT $ phase.}
		\label{Toy Model}
	\end{center}
\end{figure}
This shows that, for a given $ f_{j}(t) $, we can find a corresponding on-site potential $ \omega_j(t) $ in which the system transforms to the time-independent model. Besides, for the given time-dependant $ \omega_j(t) $, Eq. (\ref{Eq. for coupling}) gives the corresponding time-dependent couplings in which we can transform the system to a time-independent one. One can find this protocol for designing a canonical $ PT$-symmetric system.\\
To show this, we consider a toy model that can describe a couple two waveguides whose dynamic is given by Hamiltonian (\ref{time-dependant Hamiltonian}) with the following functions
\bea
&&f_1(t)=\sin(\omega t)+\epsilon_1,~~~~~~f_2(t)=\cos(\omega t)+\epsilon_2,\label{toyparameter1}\\
&&\omega_1=\dfrac{i\omega\cos(\omega t)}{\sin(\omega t)+\epsilon_1}+i\gamma,~~\omega_2=-\dfrac{i\omega\sin(\omega t)}{\cos(\omega t)+\epsilon_2}-i\gamma,\nonumber\\\label{toyparameter2}
\eea
where $ \gamma $, $ \epsilon_1 $, $ \epsilon_2 $ and $ \omega $ are real-valued constant and, we also suppose $ \nu=\nu' $. By making use of transformation $ A(t) $, the system transforms to the effective one with static $ PT $-symmetric Hamiltonian located in unbroken phase for $ \nu>\gamma $ and broken phase for $ \nu<\gamma $. These phases can be observed by looking at the quasienergies of the $ H(t) $ and mode profile in unbroken and broken phases. In fig. \ref{Toy Model}, we demonstrated the phases of the system.
\\
\textbf{Case (b).} In the second scenario, we choose $ \Gamma(\tau) $ such that the term inside bracket in Eq. (\ref{decopled differential equation}) vanishes for each effective vector state $ \zeta_{\mp} $ corresponding to stationary solution. It leads to the following choice
\be
\Gamma^{(\mp)}=\mp \tanh(\tau).
\ee
One can show that for effective on-site potential $ \Gamma^{(-)} $, the vector states $ \zeta_\pm $ are given by 
\begin{align}
	&\zeta_-=c_1^{(-)}\tau+c_2^{(-)},&&\zeta_+=c_1^{(+)}\tanh\tau+c_2^{(+)}(\tau\tanh\tau-1),\label{vector state for case2}
\end{align}
where in this case, the constants of the system are
\bea
&&c_1^{(-)}=-i\nu b,~~~~~~~c_2^{(-)}=-a,\nonumber\\
&&c_1^{(+)}=-i\nu'a+b,~c_2^{(+)}=-b.\label{constant for case2}
\eea
For $ \Gamma^{(+)} $, the vector states are same as (\ref{vector state for case2}) where the constants given by (\ref{constant for case2}) with $ a\leftrightarrow b $ and $ \nu\leftrightarrow \nu' $. If $ \omega_2(\tau)=0 $, then for the original vector state we find
\be
\Psi(\tau)_{\mp}=(\cosh\tau)^{\pm1}(f_1f_2^{-1}\zeta_{-},\zeta_{+})^T.
\ee
\textbf{Case (c).} The following relation gives the third model we study here for effective on-site potential
\be
\Gamma(\tau)=\alpha e^{i\gamma \tau}-\beta.
\ee
By introducing $ \zeta_{\mp}:=e^{-\eta_{\mp}/2}W_{\mp}(\eta_{\mp}) $, the equation (\ref{decopled differential equation}) transforms to
\be
\eta_{\mp}^2\dfrac{d^2W_{\mp}}{d\eta_{\mp}^2}+\eta_{\mp}(1-\eta_{\mp})\dfrac{dW_{\mp}}{d\eta_{\mp}}\pm\left[ \dfrac{\alpha_1^{-1}+\alpha_2}{\gamma}\mp\frac{\alpha_2^2}{\gamma^2}\right] W_{\mp}=0,
\ee
where
\begin{align}
	&\eta_{\mp}:=\mp\dfrac{2i\alpha}{\gamma}e^{i\gamma\tau},&\alpha_1:=(-i\beta+\sqrt{1-\beta^2})^{-1},\\
	&\alpha_2:=\sqrt{1-\beta^2}.
\end{align}
From here, one can obtain the general solution for $ W_\mp $ such that
\bea
&&W_-=\eta_-^{\alpha_2/\gamma}[c_1^{(-)}U(\dfrac{\alpha_1}{\gamma},1+\dfrac{2\alpha_2}{\gamma},\eta_-)+\nonumber\\
&&~~~~~~c_2^{(-)}L(\dfrac{\alpha_1^{-1}+2\alpha_2}{\gamma},\dfrac{2\alpha_2}{\gamma},\eta_-)],\\
\nonumber\\
&&W_+=\eta_+^{\alpha_2/\gamma}[c_1^{(+)}U(\dfrac{\alpha_1^{-1}}{\gamma},1+\dfrac{2\alpha_2}{\gamma},\eta_+)+\nonumber\\
&&~~~~~~c_2^{(+)}L(-\dfrac{\alpha_1^{-1}}{\gamma},\dfrac{2\alpha_2}{\gamma},\eta_+)],
\eea
where $ U $ represents confluent hypergeometric functions of the second kind, and $ L $ stands for Laguerre polynomials which satisfy the following properties \cite{zwillinger2007table}
\begin{align}
	&\frac{d}{dz}U(m_1,m_2,z)=-m_1U(m_1+1,m_2+2,z),\\
	&\frac{d}{dz}L(m_3,m_2,z)=-m_3L(m_3-1,m_2+1,z),\\
	&L(m_3,m_2,z)=\frac{(m_2+1)_{m_{3}}}{m_1!}F_1(-m_3,m_2+1,z).
\end{align}
In the third relation, $ (m_2+1)_{m_{1}} $ is Pochhammer symbol, and $ F_1 $ represents confluent hypergeometric functions of the first kind. The coefficients $ c_{1,2}^{({\mp})} $ depend on initial conditions. By following the method given in \cite{hassan2017dynamically}, we can obtain constants by using the Transfer Matrix method 
\be
\left(\begin{array}{cc}
	c_1^{(-)}   \\
	c_2^{(-)}
\end{array}\right)=z_0^{-m}e^{z_0}\left[\bM_{0}-z_1\bM_1 \right]^{-1} \left(\begin{array}{cc}
	a    \\
	b
\end{array}\right),
\label{transfer matrix relation for constants}
\ee
where the transfer matrix $ \bM_j $ is obtained as
\bea
&\bM_0=\left( \begin{array}{cc}
	\bU^{(0)}	&\bL^{(0)}  \\
	z_3\bU^{(0)}& z_3\bL^{(0)}
\end{array}\right),&\bM_2=\left( \begin{array}{cc}
	0	&0  \\
	\bU^{(1)}& \bL^{(1)}
\end{array}\right).\nonumber\\
\eea
In the above relations, we introduce
\begin{align}
	&z_0:=\dfrac{-2i\alpha}{\gamma},&&z_1:=2\alpha,\\
	&z_3:=(\dfrac{z_1m-z_1z_0}{2z_0}-i\beta),&&m:=\dfrac{\alpha_2}{\gamma},
\end{align}
and in the entries of transfer matrices, $ \bU $ and $ \bL $ stands for
\bea
&&\bU^{(j)}:=U(m_1+j,m_2+j,z=0),\\
&&\bL^{(j)}:=L(m_3-j,m_2+j,z=0). 
\eea
For $ \alpha=-\rho $ and $ \beta=-1 $, the exact solution is reduced to the one given in \cite{hassan2017dynamically}.


\begin{thebibliography}{51}
\bibitem{footnote1}
	One can consider a rigorous version of kirchhoff's laws where the time
	derivative of capacitance ($C(t)$) is taken into account. then there is an
	extra onsite potential (i.e., $\omega_1$ is no longer zero). however, in our
	approach to transforming the original hamiltonian into an effective one with
	balanced gain and loss, we consider an arbitrary onsite potentials. as a
	result, hidden pt symmetry can still be revealed in such a system with
	different broken and unbroken phases, but the location of exceptional point
	shifts.
\end{thebibliography}
\end{document}